\documentstyle[12pt, hyperref]{article}

\catcode`\@=11

\newcounter{app}

\def\app{\setcounter{equation}{0}
\def\theequation{A\arabic{app}.\arabic{equation}}\par
   \addvspace{4ex}
   \@afterindentfalse
  \secdef\@app\@dapp}
\newcommand\@app{\@startsection {app}{1}{0ex}%
                                   {-3.5ex \@plus -1ex \@minus -.2ex}%
                                   {2.3ex \@plus.2ex}%
                                   {\normalfont\Large\bf}}

\def\@dapp#1{%
{\parindent \z@ \raggedright  \bf #1}\par\nobreak}
\def\l@app#1#2{\ifnum \c@tocdepth >\z@
    \addpenalty\@secpenalty
    \addvspace{1.0em \@plus\p@}%
    \setlength\@tempdima{8.5em}%
    \begingroup
      \parindent \z@ \rightskip \@pnumwidth
      \parfillskip -\@pnumwidth
      \leavevmode \bfseries
      \advance\leftskip\@tempdima
      \hskip -\leftskip
      #1\nobreak\hfil \nobreak\hb@xt@\@pnumwidth{\hss #2}\par
    \endgroup\fi}
\catcode`\@=12

\small  \oddsidemargin -10mm  \evensidemargin 5mm  \topmargin -10mm
\def\be{\begin{equation}}
\def\ee{\end{equation}}

\newtheorem{theorem}{Theorem}
\newtheorem{prop}{Proposition}
\newtheorem{remark}{Remark}
\newtheorem{consequence}{Consequence}

\def\bprop{\begin{prop}}
\def\eprop{\end{prop}}
\def\bremark{\begin{remark}}
\def\eremark{\end{remark}}
\begin{document}
\author{A. Yu. Orlov\thanks{Oceanology Institute, Nahimovskii
prospect 36, Moscow, Russia, email address:
orlovs@wave.sio.rssi.ru}}
\title{New solvable Matrix Integrals}
\date{}
\maketitle

\begin{abstract}
We generalize the Harish-Chandra-Itzykson-Zuber and certain other
integrals (Gross-Witten integral and integrals over complex
matrices) using the notion of tau function of matrix argument. In
this case one can reduce multi-matrix integrals to the integral
over eigenvalues, which in turn is a certain tau function. We also
consider a generalization of the Kontsevich integral.

\end{abstract}

\section{Introduction}

The study of different models of random matrices is one of the
most interesting topic in mathematical physics (see
\cite{Mehta},\cite{TW},\cite{ZJZ},\cite{Moe} for a review). Models
of random matrices are applied to various topics starting with
quantum field theory and condensed matter physics ending with
combinatorics, algebraic topology and number theory. The purpose
of the paper is to generalize some known solvable matrix
integrals. By the {\em solvable matrix integrals} we mean those
which can be presented as tau functions and which can be evaluated
by the method of orthogonal polynomials. Let us mark that in
papers \cite{Kaz1},\cite{Kaz2} by solvability of matrix model the
author means a weaker condition - a model considered to be
solvable if it admits a reduction of number of integrations (from
the order ${\sim}N^2$ to the order $N$) via a character expansion
method without any relation to the tau function topic.

Today it is not surprising that the notion of tau function,
introduced by Sato school (see a list of references in \cite{JM}),
is widely used in statistical physics and quantum field theory as
a partition function of different models (as an example see the
review \cite{M}). The special class of tau functions, which we
shall exploit in the present paper and which we call  tau
functions of hypergeometric type, is not an exclusion. Tau
functions of hypergeometric type are parameterized by a function
$r$ on the lattice and are denoted by $\tau_r$. (The name
hypergeometric tau function \cite{pd22} is due to the fact that
the simplest specialization, $r$ is rational, yields the
hypergeometric function of matrix argument \cite{GR},\cite{V}.)
This is a type of tau functions one may meet in many different
problems, and special examples of these tau functions one can find
in \cite{NTT},\cite{T},\cite{O2}, \cite{Nekrasov}, all these
examples one may obtain by a specialization of $r$. The most of
well known matrix integrals actually belong to this class of
tau-functions \cite{1},\cite{HO'},\cite{Or}. (To be precise we
deal with asymptotic expansion of matrix integrals and tau
functions).

In the present paper we show that the integrals of tau functions
of hypergeometric type of matrix argument are again tau functions
of hypergeometric type. We notice that all known matrix integrals
such as one-matrix model \cite{K}, Itzykson-Zuber integral
\cite{HC},\cite{IZ}, two-matrix matrix model \cite{IZ},
Gross-Witten one plaquette model \cite{GW} may be considered as
integrals of the simplest tau function (so-called vacuum
 tau function). The determinant
representation of the tau functions of hypergeometric type allows
to obtain generalizations of solvable multi-matrix models and a
certain generalization of Kontsevich integral.

In the last part of this introduction we  review some things from
soliton theory.

{\bf Soliton theory}. KP hierarchy of integrable equations
\cite{ZSh},\cite{JM}, which is the most popular example,  consists
of semi-infinite set of nonlinear partial differential equations
\begin{equation}\label{KPh}
\partial_{t_m}u=K_m[u] \ ,\quad m=1,2,\dots \ ,
\end{equation}
which are commuting flows: $
\left[\partial_{t_k},\partial_{t_m}\right]u=0 $. The first
nontrivial one is Kadomtsev-Petviashvili equation
\begin{equation}\label{KP}
\partial_{t_3}u=\frac 14 \partial_{t_1}^3 u+ \frac 34
\partial_{t_1}^{-1}\partial_{t_2}^2u+\frac 34 \partial_{t_1} u^2 \ ,
\end{equation}
which originally served in plasma physics \cite{ZSh}, now plays a
very important role both, in physics  and in mathematics. Another
very important equation is the equation of two-dimensional Toda
lattice (TL) carefully studied in \cite{UT}
\begin{equation}\label{Toda}
\partial_{t_1}\partial_{t^*_1}\phi_n=
e^{\phi_{n-1}-\phi_{n}}- e^{\phi_{n}-\phi_{n+1}}
\end{equation}
This equation gives rise to TL hierarchy which contains
derivatives with respect to the higher times $t_1,t_2,\dots $ and
$t_1^*,t_2^*,\dots \ $.

The key point of the soliton theory is the notion of tau function,
introduced by Sato (for KP tau-function see \cite{JM}). The tau
function is a sort of a potential which gives rise both to TL
hierarchy and KP hierarchy. It depends on two semi-infinite sets
of higher times $t_1,t_2,\dots$ and $t_1^*,t_2^*,\dots$, and on
discrete variable $n$: $\tau=\tau(n,{\bf t},{\bf t^*})$. More
explicitly we have \cite{JM},\cite{UT}:
\begin{equation}\label{tauuphi}
\quad u=2\partial_{t_1}^2\log \tau(n,{\bf t},{\bf t^*}) \ ,\quad
\phi_n({\bf t},{\bf t}^*)=-\log \frac{\tau(n+1,{\bf t},{\bf
t}^*)}{\tau(n,{\bf t},{\bf t}^*)}
\end{equation}

In the soliton theory so-called Hirota-Miwa  variables ${\bf
x},{\bf y}$, which are related to the higher times as
\begin{equation}\label{HMxy}
mt_m=\sum_i x_i^m \ ,\quad mt_m^*=\sum_i y_i^m \ ,
\end{equation}
are well-known. Any tau function is a symmetric function in
Hirota-Miwa variables, the higher times $mt_m,mt_m^*$ are
so-called \cite{Mac} power sums.

It is known fact that a typical tau function may be presented in
the form of double series over partitions in Schur functions
\cite{Tinit}:
\begin{equation}\label{tauschurschur}
\tau(n,{\bf t},{\bf t^*})=\sum_{\lambda,\mu}K_{\lambda
\mu}s_\lambda({\bf t})s_\mu ({\bf t^*}) \ ,
\end{equation}
where the coefficients $K_{\lambda \mu }$  solve special bilinear
equations \cite{JM}.

{\bf $\tau$ functions of hypergeometric type}. Let us consider a
function $r$ which depends on a single variable $n$, the $n$ is an
integer. We suppose $r(n)$ to be finite for integer $n$. Given
partition $\lambda$, we define
\begin{equation}\label{rlambda}
r_\lambda(x)=\prod_{i,j\in \lambda}r(x+j-i)
\end{equation}
Namely, $r_\lambda(n)$ is a product of $r$ over all nodes of the
Young diagram of the partition $\lambda$ where argument of $r$ is
defined by entries $i,j$ of a node \cite{Mac}. The value of $j-i$
is zero on the main diagonal; the value $j-i$ is called the
content of the node. For zero partition one puts $r_0 \equiv 1$.
It was shown \cite{tmf} that
\begin{equation}\label{exex}
\tau_r(n,{\bf t},{\bf t}^*)=\sum_\lambda r_\lambda (n)
s_\lambda({\bf t})s_\lambda({\bf t^*})
\end{equation}
(where the sum is going over all partitions including zero one) is
a TL and a KP tau function, which we call the tau function of
hypergeometric type. The Schur functions $s_\lambda({\bf
t}),s_\lambda({\bf t^*})$ are defined with the help of
\begin{equation}\label{Schurtt*}
s_\lambda({\bf t})=\det h_{\lambda_i-i+j}({\bf t})_{1\le i,j\le
l(\lambda) } \ ,\quad \exp\sum_{m=1}^\infty
z^mt_m=\sum_{k=1}^\infty z^kh_k({\bf t})
\end{equation}
Here $h_k({\bf t})$ is called the elementary Schur function, or,
the same, the complete symmetric function, see \cite{Mac}. We
should say more about the case ${\bf t}={\bf t}({\bf x}^n)$. Mark
that $s_\lambda({\bf t}({\bf x}^n))=0$ if the length of the
partition $l(\lambda)$ exceeds $n$, therefore, in this case, the
sum over $\lambda$ (\ref{exex}) is restricted by $ l(\lambda) \le
n$. Due to (\ref{rlambda}), it means that $r(k),k\le 0$  are not
involved to the series (\ref{exex}) if ${\bf t}={\bf t}({\bf
x}^n)$.

If ${\bf x}^n=(x_1,\dots,x_n)$ are eigenvalues of a $ n$ by $n$
matrix $X$, and the variables ${\bf t}$ and ${\bf x}^n$ are
related by (\ref{HMxy}), we write $s_\lambda(X):=s_\lambda({\bf
t}({\bf x}^n))$.
 It is suitable to introduce the notion of {\em tau function of
matrix argument} by analogy with the hypergeometric function of
matrix argument (see (\ref{taushurF2}),(\ref{taushurmathcalF2})
below):
\begin{equation}\label{taumatr}
\tau\left(n,X,{\bf t}^*\right):=\tau\left(n,{\bf t}({\bf
x}^n),{\bf t}^* \right) \ ,\quad
\tau\left(n,X,Y\right):=\tau\left(n,{\bf t}({\bf x}^n),{\bf
t}^*({\bf y}^n) \right) \ ,
\end{equation}
where Hirota-Miwa variables $x_1,\dots ,x_n$ are just eigenvalues
of a matrix $X$. We shall use large letters for this matrix
argument. As in the case of the hypergeometric functions the tau
function by definition depends only on eigenvalues of the matrix.

\section{ Useful formulae}  We shall exploit the
following formulae for integrations of Schur functions over the
unitary group \cite{Mac}. Let $d_*U$ be the normalized Haar
measure on the $U(n)$ and let $\delta_{\mu,\lambda}$ be the
Kronecker symbol. Then
\begin{equation}\label{sAUBU^+}
\int_{U(n)}s_\lambda(AUBU^{-1})d_*U=
\frac{s_\lambda(A)s_\lambda(B)}{s_\lambda(I_n)} \ ,
\end{equation}
\begin{equation}\label{sAUU^+B}
\int_{U(n)}s_\mu(AU)s_\lambda(U^{-1}B)d_*U=
\frac{s_\lambda(AB)}{s_\lambda(I_n)}\delta_{\mu,\lambda}
\end{equation}
Let ${\bf t}_{\infty}=(1,0,0,0,\dots)$. For integration over
complex matrices $Z$ there are the formulae \cite{Mac}
\begin{equation}\label{sAZBZ^+}
\int_{C^{n^2}} s_\lambda(AZBZ^+)e^{-\textrm{Tr}
ZZ^+}\prod_{i,j=1}^n d^2Z=
\frac{s_\lambda(A)s_\lambda(B)}{s_\lambda({\bf t}_{\infty})}
\end{equation}
and
\begin{equation}\label{sAZZ^+B}
\int_{C^{n^2}} s_\mu(AZ)s_\lambda(Z^+B) e^{-\textrm{Tr}
ZZ^+}\prod_{i,j=1}^nd^2Z= \frac{s_\lambda(AB)}{s_\lambda({\bf
t}_{\infty})}\delta_{\mu,\lambda}
\end{equation}
Throughout the text  $d^2Z=\pi^{-n^2}\prod_{i,j=1}^n{d\Re
Z_{ij}d\Im Z_{ij}}$; at sequel we shall omit $C^{n^2}$. Then we
need \cite{Mac}
\begin{equation}\label{Hs}
s_\lambda({\bf t}_{\infty})=\frac{1}{H_\lambda} \ ,\quad
H_\lambda=\prod_{i,j}(\lambda_i+\lambda_j'-i-j+1) \ ,
\end{equation}
where $H_\lambda$ is the hook product. The {\em Pochhammer's
symbol related to a partition
$\lambda=(\lambda_1,\dots,\lambda_k)$} is the following product of
the Pochhammer's symbols
$(a)_\lambda=(a)_{\lambda_1}(a-1)_{\lambda_2}
\cdots(a-k+1)_{\lambda_k}$, $(a)_{\lambda_i}=\Gamma(a+\lambda_i)/
\Gamma(a)$. Let ${\bf t}(a)=(\frac a1 ,\frac a2,\frac a3,\dots)$.
Then
\begin{equation}\label{sIn}
(a)_\lambda=H_\lambda s_\lambda({\bf t}(a))=\frac{s_\lambda({\bf
t}(a))}{s_\lambda({\bf t}_{\infty})} \ ,\quad
(n)_\lambda=H_\lambda
s_\lambda(I_n)=\frac{s_\lambda(I_n)}{s_\lambda({\bf t}_{\infty})}
\end{equation}

In addition we have a simple relation (bosonic (l.h.s) and
fermionic (r.h.s) representation of the vacuum TL tau function,
since we did not find it in literature we proved it in \cite{HO}):
\begin{equation}\label{epp2}
\exp \sum_{m=1}^\infty mt_mt_m^*= \sum_{\lambda}s_\lambda({\bf
t})s_\lambda({\bf t}^*) \ ,
\end{equation}
which is a generalized version  of {\em Cauchy-Littlewood identity
\cite{Mac}}.

Let $r(k)=q^{k}$.  One gets tau function used in \cite{OP} as a
generating function for double Hurwitz numbers.

 Let
$r(k)=\prod_{i=1}^p(k+a_i)\prod_{i=1}^s(k+b_i)^{-1}$, and let
$l(\lambda)$ be the number of non-vanishing parts of $\lambda$.
Let us write down the formulae for the {\em hypergeometric
function of a matrix argument $X$ \cite{V}},\cite{tmf} (throughout
the paper all sums include zero partition)
\begin{eqnarray}\label{taushurF2}
{}_p{F}_s\left.\left(a_1+M,\dots ,a_p+M\atop b_1+M, \cdots
,b_s+M\right| X\right)=
\tau_{r}(M,{\bf t}({\bf x}^n),{\bf t}_\infty)=\nonumber\\
\sum_{\lambda\atop l(\lambda)\le n } \frac{\prod_{k=1}^p
s_{\lambda}({\bf t}(a_k+M))} {\prod_{k=1}^s s_{\lambda}({\bf
t}(b_k+M))} \left(s_{\lambda}({\bf t}_\infty)\right)^{s-p+1}
s_{\lambda}(X)
\end{eqnarray}
Examples:
\begin{equation}\label{examples}
{}_0{F}_0(X)=e^{ \textrm{Tr} X}\ ,\quad
{}_1{F}_0(a|X)=e^{\sum_{m=1}^\infty \frac am \textrm{Tr}
X^m}=\det(1-X)^{-a}
\end{equation}

 Now let
$r(k)=
{\prod_{i=1}^p(k+a_i)}{(k+n-M)^{-1}}{\prod_{i=1}^s(k+b_i)^{-1}}$.
The {\em hypergeometric functions of two matrix arguments $X,Y$ is
\cite{V}},\cite{tmf}
\begin{eqnarray}\label{taushurmathcalF2}
{}_p{\mathcal{F}}_s\left.\left(a_1+M,\dots ,a_p+M\atop b_1+M,
\cdots ,b_s+M\right| X,Y\right)=
\tau_{r}(M,{\bf t}({\bf x}^n),{\bf t}^*({\bf y}^n))=\nonumber\\
\sum_{\lambda\atop l(\lambda)\le n } \frac{\prod_{k=1}^p
s_{\lambda}({\bf t}(a_k+M))} {\prod_{k=1}^s s_{\lambda}({\bf
t}(b_k+M))} \left(s_{\lambda}({\bf t}_\infty)\right)^{s-p+1}\frac
{s_{\lambda}(X)s_{\lambda}(Y)} {s_{\lambda}(I_n)}
\end{eqnarray}
Here ${\bf x}^n=(x_1,\dots,x_n),{\bf y}^n=(y_1,\dots,y_n)$ are
eigenvalues of the matrices $X,Y$. Matrices $X,Y$ are supposed to
be diagonalizable (i.e. {\em normal}) matrices.

 Taking $b_1=-c+\epsilon$ where $c\ge -n$ is an
 integer,  we obtain \cite{Or}
\begin{equation}\label{pFqcnegativematr}
\lim_{\epsilon \to 0}\ {}_pF_q\left.\left(a_1,\dots,a_p \atop
-c+\epsilon, b_2,\dots , b_q \right| X
\right)\frac{(-c+\epsilon)_\sigma,(b_2)_\sigma \cdots (b_q)_\sigma
}{(a_1)_\sigma \cdots (a_p)_\sigma} (n)_\sigma
\end{equation}
\begin{equation}\label{pFqcnegativematr'}
= \det X^{n+c}
 {}_pF_q\left.\left(a_1+n+c,\dots ,a_p+n+c
 \atop 2n+c,b_2+n+c,\dots ,b_q+n+c
\right| X \right) \ ,
\end{equation}
where $\sigma$ is the partition $(n+c,\dots ,n+c)$ with the length
$ l(\sigma)=n $. Meanwhile
\begin{equation}\label{pFqXYcnegativematr}
\lim_{\epsilon \to 0}\ {}_p{\mathcal{F}}_q\left.\left(a_1,\dots
,a_p\atop -c+\epsilon , b_2, \cdots ,b_q\right|
X,Y\right)\frac{(-c+\epsilon)_\sigma (b_2)_\sigma \cdots
(b_q)_\sigma }{(a_1)_\sigma \cdots (a_p)_\sigma} (n)_\sigma
\end{equation}
\begin{equation}\label{pFqXYcnegativematr'}
=\det X^{n+c}\det Y^{n+c}{}_p{\mathcal{F}}_q
\left.\left(a_1+n+c,\dots ,a_p+n+c
 \atop 2n+c,b_2+n+c,\dots ,b_q+n+c
\right| X , Y \right) \
\end{equation}
In particular
\begin{equation}\label{1F1cnegativematr}
\lim_{\epsilon \to 0}\ {}_1F_1\left.\left(n \atop -c+\epsilon
\right| X \right)(-c+\epsilon)_\sigma={}_0F_0\left( X \right)\det
X^{n+c}=e^{ \textrm{Tr} X}\det X^{n+c} \ ,
\end{equation}
\begin{equation}\label{2F1cnegativematr}
\lim_{\epsilon \to 0}\ {}_2F_1\left.\left(a,n \atop -c+\epsilon
\right| X \right)\frac{(-c+\epsilon)_\sigma}{(a)_\sigma}=\det
X^{n+c}{}_1F_0\left(a+n+c | X \right)=\det
X^{n+c}\det(1-X)^{-a-n-c}
\end{equation}
These relations are particular cases of the limit of $\tau_r$,
where we put $r$ to be singular: $r$ is evaluated at the vicinity
of zero as $r(\epsilon)=\frac 1\epsilon+O(\epsilon)$. Below $c\ge
-n$ is an integer and $\sigma$ is a partition $(n+c,\dots,n+c)$ of
the length $l(\sigma)=n$. Then we have \cite{Or}
\begin{equation}\label{taursingular1detX}
\lim_{\epsilon \to 0}\ \frac{\tau_r(-c+\epsilon,{\bf t}({\bf
x}^n),{\bf t }_\infty)}{r_{\sigma}(-c+\epsilon)}=\frac{\det
X^{n+c}}{(n)_\sigma } \tau_{\tilde r}(n,{\bf t}({\bf x}^n),{\bf
t}_\infty)\ ,\quad \tilde r(k)=r(k)\frac{k}{k+n+c} \ ,
\end{equation}
\begin{equation}\label{taursingular1detXY}
\lim_{\epsilon \to 0}\ \frac{\tau_r(-c+\epsilon,{\bf t}({\bf
x}^n),{\bf t^* }({\bf y}^n))}{r_{\sigma}(-c+\epsilon)}=\det
X^{n+c}\det Y^{n+c} \tau_r(n,{\bf t}({\bf x}^n),{\bf t^* }({\bf
y}^n))
\end{equation}

 Let us mark also the {\bf determinant representations of
(\ref{taushurF2}) and of (\ref{taushurmathcalF2})} which are of
importance in applications of the method of orthogonal polynomials
to the new matrix models, which we shall consider below.

The determinant representations have forms (the reader may find
details in \cite{1},\cite{tmf}):
\begin{equation}\label{det1'}
\tau_r(M,{\bf t}({\bf x}^n),{{\bf t}^*})=\frac{ \det \
x_i^{n-k}\tau_r(M-k+1,{\bf t}(x_i),{{\bf t}^*})\ |\
_{i,k=1}^n}{\Delta({\bf x}^n)} \ ,
\end{equation}
(which in particular yields the determinant representation of
 ${}_p{F}_s$ of (\ref{taushurF2})) and
\begin{equation}\label{det2'}
 \tau_r(M,{\bf
t}({\bf x}^n),{\bf t^*}({\bf y}^n))= c_n(M)\frac{\det \
\tau_r(M-n+1,x_i,y_j)\ |\ _{i,k=1}^n} {\Delta({\bf
x}^n)\Delta({\bf y}^n)} \ ,
\end{equation}
(which  yields the determinant representation of
${}_p{\mathcal{F}}_s$, see (\ref{detF2}) below), where
\begin{equation}\label{vandermond}
\Delta({\bf x}^n)=\prod_{1\le i<j\le n} (x_i-x_j)=\det (x_i^{n-k})
\ ,
\end{equation}
\begin{equation}\label{cn}
c_1=1 \ ,\qquad c_n(M)=\prod_{k=1}^{n-1}\left(
r(M-n+k)\right)^{k-n}  \ ,\quad n>1 \ ,
\end{equation}
\begin{equation}\label{tau(1)}
\tau_r(M-n+1,x_i,y_j)=1+r(M-n+1)x_iy_j+
r(M-n+1)r(M-n+2)x_i^2y_j^2+\cdots \ ,
\end{equation}
where $c_n^{-1}(M)\Delta({\bf x}^n)\Delta({\bf y}^n)$ is supposed
to be non-vanishing; otherwise the formula (\ref{det2'}) should be
modified, see \cite{Or}. Most often we need the cases $M=0,n$.
Examples of (\ref{det1'}),(\ref{det2'}):
\begin{equation}\label{detF1}
{_pF}_s\left.\left(a_1,\dots ,a_p\atop b_1, \cdots ,b_s\right|
X\right)=\frac{1}{\Delta({\bf x}^n)} \det x_i^{n-k}
{}_p{F}_s\left.\left(a_1-k+1,\dots ,a_p-k+1\atop b_1-k+1, \cdots
,b_s-k+1\right| x_i\right)|\ _{i,k=1}^n \ ,
\end{equation}
\begin{equation}\label{detF2}
{}_p{\mathcal{F}}_s\left.\left(a_1+n,\dots ,a_p+n\atop b_1+n,
\cdots ,b_s+n\right| X,Y\right)=\frac{c_n(n)}{\Delta({\bf
x}^n)\Delta({\bf y}^n)} \det {}_p{F}_s\left.\left(a_1+1,\dots
,a_p+1\atop b_1+1, \cdots ,b_s+1\right| x_iy_j\right)|\ _{i,j=1}^n
\end{equation}
If $r(0)=0$ there is the following formula for the open Toda
lattice (for instance see \cite{GMMMO},\cite{1}):
\begin{equation}\label{D3detzeroesr}
\tau_r(n,{\bf t},{\bf t^*})=c_n\det \left(
\frac{\partial^{i+k-2}}{\partial t_1^{i-1}\partial
{t_1^*}^{k-1}}\tau_r(1,{\bf t},{\bf t^*})\right)|\ _{i,k=1}^n \ ,
\end{equation}
\begin{equation}\label{D3cn}
c_n=\prod_{k=1}^{n-1}\left( r(k)\right)^{k-n}
\end{equation}

At last let us notice that $\tau_r(n,{\bf t},{\bf t}^*)$ enjoys
the following symmetry relations:
\begin{equation}\label{symmtau1}
\tau_r(n,{\bf t},{\bf t}^*)=\tau_r(n,{\bf t}^*,{\bf t}) \ ,\quad
\tau_r(n+m,{\bf t},{\bf t}^*)=\tau_{\tilde r}(n,{\bf t}^*,{\bf t})
\ ,
\end{equation}
\begin{equation}\label{symmtau}
\tau_r(n,{\bf t}_\infty,{\bf t}^*)=\tau_{r_1}(n,{\bf t}(a),{\bf
t}^*)=\tau_{r_2}(n,I_n,{\bf t}^*) \ ,
\end{equation}
where $I_n$ is $n$ by $n$ unit matrix, and where ${\tilde
r}(k)=r(k+m),\ r_1(k)=(k+a)^{-1}r(k),\ r_2(k)=(k+n)^{-1}r(k)$, the
 relations (\ref{symmtau}) are derived from (\ref{sIn}), for details see
\cite{tmf}.

\section{Tau functions of a matrix argument and angle
integration.
 The integration of tau functions over complex matrices}

\begin{theorem} Let $x_i$ and $y_i$, where $i=1,\dots , n$, are
eigenvalues of matrices $X$ and $Y$ respectively. If for $n>1$
\begin{equation}\label{cn}
c_n(n)=\prod_{k=1}^{n-1}\left( r(k)\right)^{k-n} \neq 0 \ ,
\end{equation}
then there is the following generalization of HCIZ integral
\begin{equation}\label{GHCIZa}
  \int_{U(n)}  \tau_r\left(n,XUYU^+,I_n\right)
  d_*U= \tau_r\left(n,X,Y\right)= c_n(n)\frac{\det
\left(\tau_r(1,x_i,y_j)\right)_{i,j=1,\dots ,n}} {\Delta({\bf
x}^n)\Delta({\bf y}^n)} \ ,
\end{equation}
where ${\Delta({\bf x}^n),\Delta({\bf y}^n)}$ are the Vandermond
determinants.
\end{theorem}
The proof follows directly from
(\ref{exex}),(\ref{sAUBU^+}),(\ref{det2'}).
 In particular we have (see (\ref{taushurF2}),(\ref{taushurmathcalF2}))
\begin{equation}\label{GHCIZaF}
  \int_{U(n)} {}_p{F}_s\left.\left(a_1,\dots ,a_p\atop b_1,
\cdots ,b_s\right| XUYU^+\right)d_*U
={}_p{\mathcal{F}}_s\left.\left(a_1,\dots ,a_p\atop b_1, \cdots
,b_s\right| X,Y\right)
\end{equation}

Let us consider the simplest examples of (\ref{GHCIZaF}), which,
at the same time, are the solvable cases of
\begin{equation}\label{eAUBU+}
\int_{U(n)} e^{\sum_{m=1}^\infty t_m
\textrm{Tr}\left(XUYU^+\right)^m}d_*U =\sum_{\lambda,l(\lambda)\le
n} \frac {s_\lambda({\bf t})
s_\lambda(X)s_\lambda(Y)}{s_\lambda(I_n)}
\end{equation}
Similar series for matrix integrals were considered in the papers
\cite{Kaz1},\cite{Kaz2} without connections with classical
integrable systems.

(1) HCIZ integral,
\cite{HC},\cite{IZ},\cite{Mehta},\cite{ZJ},\cite{ZJZ}, which we
obtain choosing $p=s=0$
\begin{equation}\label{C1}
\int_{U(n)} e^{ \textrm{Tr}\left(XUYU^+\right)}d_*U
=\sum_{\lambda,l(\lambda)\le n} \frac
{s_\lambda(X)s_\lambda(Y)}{H_\lambda s_\lambda( I_n)}
={_0{\mathcal{F}}}_0\left({{{X}}} \ ,
{{{Y}}}\right)=c_n\frac{\det( e^{x_iy_j})}{\Delta(x)\Delta(y)} \ ,
\end{equation}
\begin{equation}\label{C1r}
c_1=1 \ ,\quad c_n=c_n(n)=\prod_{k=1}^{n-1}\left(
r(k)\right)^{k-n}= \prod_{k=1}^{n-1}\left(\frac{1}{k}\right)^{k-n}
\  ,\ n>1 \ ,\qquad r(k)=\frac{1}{k}
\end{equation}

(2) Choosing $p=s+1=1$ we get
\begin{equation}\label{C2}
\int_{U(n)} \det\left(1- XUYU^+\right)^{-a} d_*U
=\sum_{\lambda,l(\lambda)\le n} (a)_\lambda \frac{
{s_\lambda(X)s_\lambda(Y)}}{H_\lambda s_\lambda
(I_n)}=\tau_r(n,X,Y)
\end{equation}
\begin{equation}\label{C2det}
={_1{\mathcal{F}}}_0\left(a ;{{{X}}} \ ,
{{{Y}}}\right)=c_n\frac{\det(1-x_iy_j)^{n-1-a}}{\Delta({\bf
x}^n)\Delta({\bf y}^n)} \  ,
\end{equation}
\begin{equation}\label{C2r}
c_n=c_n(n)=\prod_{k=1}^{n-1}\left( r(k)\right)^{k-n} \ ,\quad n>1
\ ,\qquad r(k)=\frac{a-n+k}{k}
\end{equation}

\quad

Now let us consider integrals over complex matrices.  Different
integrals over complex matrices were considered in the papers of
I.Kostov, see \cite{Kos1},\cite{Kos2} as an example.

\begin{theorem} Under the conditions of the Theorem 1 we have
\begin{equation}\label{complexgen'}
 \int  \tau_r\left(n,{\bf t}_\infty,XZYZ^+\right)
  e^{-\textrm{Tr}ZZ^+}d^2Z =\tau_r\left(n,X,Y\right)
= c_n(n)\frac{\det \left(\tau_r(1,x_i,y_j)\right)_{i,j=1,\dots
,n}} {\Delta({\bf x}^n)\Delta({\bf y}^n)} \,
\end{equation}
where $d^2Z=\pi^{-n^2}\prod_{i,j=1}^nd\Re Z_{ij}d\Im Z_{ij}$. In
particular we have
\begin{equation}\label{complexgen'F}
  \int {}_p{F}_s\left.\left(a_1,\dots ,a_p\atop b_1,
\cdots ,b_s\right|XZYZ^+ \right)e^{-\textrm{Tr}ZZ^+}d^2Z
={}_{p+1}{\mathcal{F}}_s\left.\left(n, a_1,\dots ,a_p\atop b_1,
\cdots ,b_s\right| X,Y\right)
\end{equation}

\end{theorem}
The proof follows from
(\ref{exex}),(\ref{sAZBZ^+}),(\ref{det2'}),(\ref{taushurF2}),
(\ref{taushurmathcalF2}). Examples of solvable models provided by
(\ref{complexgen'F}):

(1) $p=s=0$, then
\begin{equation}\label{A1}
\int e^{\sum_{m=1}^\infty
\textrm{Tr}\left(XZYZ^+\right)}e^{-\textrm{Tr}ZZ^+}d^2Z
=\sum_{\lambda,l(\lambda)\le n}
s_\lambda(X)s_\lambda(Y)=\prod_{i,j}^n(1-x_iy_j)^{-1}
={_1{\mathcal{F}}}_0\left(n|{{{X}}}, {{{Y}}}\right)
\end{equation}

(2) $p=s+1=1$:
\begin{equation}\label{A2}
\int e^{\sum_{m=1}^\infty \frac{a}{m}
\textrm{Tr}\left(XZYZ^+\right)^m}e^{-\textrm{Tr}ZZ^+}d^2Z =\int
\det\left(1-\frac 12 XZYZ^+\right)^{-a} e^{-\textrm{Tr}ZZ^+}d^2Z
\end{equation}
\begin{equation}\label{A2'}
=\sum_{\lambda,l(\lambda)\le n} (a)_\lambda
{s_\lambda(X)s_\lambda(Y)}={_2{\mathcal{F}}}_0\left(a,n|{{{X}}},
{{{Y}}}\right) \ ,
\end{equation}
which in general is a divergent series (until $-a$ is a
non-negative integer).

\quad

 Next theorem is a generalization of Gross-Witten integral:
\begin{theorem} Let $z_i,i=1,\dots,n$ be eigenvalues of the matrix $XY$
\begin{equation}\label{tauGW}
\int_{U(n)}\tau_{ r}\left( n,{\bf t}, X U\right) \tau_{\tilde
r}\left( n, U^{-1}Y ,I_n\right)d_*U= \tau_{r\tilde r}\left( n,
{\bf t} ,XY\right)
\end{equation}
\begin{equation}\label{tauGW'}
=\frac{ \det\left(z_i^{n-k}\tau_{r\tilde r}(n-k+1,{{\bf t}},{\bf
t}^*(z_i)) \right)_{i,k=1,\dots ,n}}{\Delta({\bf z}^n)} \ ,
\end{equation}
where $\Delta({\bf z}^n)$ is the Vandermond determinant. Also
\begin{equation}\label{tauGW2}
\int_{U(n)}\tau_{ r}\left( n,{\bf t}, X U\right) \tau_{\tilde
r}\left( n, U^{-1}X^{-1} ,{\bf t}^*\right)d_*U= \tau_{r_1}\left(
n, {\bf t} ,{\bf t}^*\right)\ ,
\end{equation}
where $r_1$ is the following step function:
\begin{equation}\label{r1}
r_1(k)=r(k) {\tilde r}(k)\ ,\ k>0 \ , \quad r(k)=0 \ ,\ k\le 0
\end{equation}
\end{theorem}
For the proof we use (\ref{sAUU^+B}),(\ref{det1'}). The simplest
example of (\ref{tauGW2}), $X=1$ , $r={\tilde r}=1$ is the
well-known model of unitary matrices, see \cite{ZKMMO} about this
model. We obtain
\begin{equation}\label{D3}
\int_{U(n)}e^{ \sum_i t_i\textrm{Tr}U^i +\sum_i
t^*_i\textrm{Tr}U^{-i}}d_*U=\sum_{\lambda,l(\lambda)\le n}
s_\lambda({\bf t}) s_\lambda ({\bf t}^*)=\tau_r(n,{\bf t},{\bf
t}^*)
\end{equation}
In (\ref{D3}) $r$ is the following step function: $r(k)=1  ,\ k>0
\ ; \ r(k)=0  ,\ k\le 0$.

Due to the restriction $l(\lambda)\le n$ it is not the l.h.s of
Cauchy-Littlewood formula (\ref{epp2}).  Let us mark that the
right hand side of (\ref{D3}) is the subject of the so-called
Gessel theorem  if one take ${\bf t}={\bf t}({\bf x}^m),\quad {\bf
t}^*={\bf t}^*({\bf y}^m),\quad n<m$. This case was considered in
\cite{Moe}. There exist different determinant representations of
(\ref{D3}), which are due to (\ref{det1'}),  a modification of
(\ref{det2'}) and due to (\ref{D3detzeroesr}), see \cite{Or}.

An example of (\ref{tauGW}) is
\begin{equation}\label{D1}
\int_{U(n)} e^{ \sum_m t_m\textrm{Tr}(X U)^m}
 \det\left(1-U^{-1}Y\right)^{-b}d_*U
=\sum_{\lambda,l(\lambda)\le n} (b)_\lambda \frac{s_\lambda({\bf
t})s_\lambda(XY)}{(n)_\lambda}
\end{equation}
\begin{equation}\label{D1det}
=\frac{ \det\left(z_i^{n-k}\tau_r(n-k+1,{\bf t}(z_i),{{\bf t}})
\right)_{i,k=1}^{n}}{\Delta({\bf z}^n)} \ ,\quad \tau_r(n-k+1,{\bf
t}(z_i),{{\bf t}})=\sum_{j=0}^\infty
\frac{(b-k+1)_j}{(n-k+1)_j}z_i^jh_j({\bf t})
\end{equation}
where $z_i,i=1,\dots ,n$ are eigenvalues of the matrix $XY$. Here
$h_k({\bf t})$ is the elementary Schur function, see
(\ref{Schurtt*}).

\begin{consequence} Let  $A, X, Y$ are $n$ by $n$ normal
matrices.  For integer $m\ge 0$ consider the partition
$\sigma=(m,\dots , m)$ of the length $l(\sigma)=n$. We have
\begin{equation}\label{tauGWcon}
\int_{U(n)}\tau_{ r}\left( n,A, X U\right) \tau_{\tilde r}\left(
n, U^{-1}Y ,I_n\right)\det U^{-m}d_*U= \det A^{m}\det
X^{m}r_\sigma(n)\tau_{r_1}\left( n, A ,XY\right) \ ,
\end{equation}
where $r_1(k)=r(k+m){\tilde r}(k)$. Also
\begin{equation}\label{tauGWcon'}
\int_{U(n)}\tau_{ r}\left( n,A, X U\right) \tau_{\tilde r}\left(
n, U^{-1}Y ,I_n\right)\det U^{m}d_*U= \det Y^{m}{\tilde
r}_\sigma(n)\tau_{r_1}\left( n, A ,XY\right) \ ,
\end{equation}
where $r_1(k)=r(k){\tilde r}(k+m)$. In particular
\begin{equation}\label{FF1-}
\int_{U(n)} \det U^{-m}{}_p{\mathcal{F}}_s\left.\left(a_1,\dots
,a_p\atop b_1, \cdots ,b_s\right| A, XU\right){}_{\tilde
p}{F}_{\tilde s}\left.\left({\tilde a_1},\dots ,{\tilde a_{\tilde
p}}\atop {\tilde b_1}, \cdots ,{\tilde b_{\tilde s}}\right|
U^{-1}Y\right)d_*U
\end{equation}
\begin{equation}\label{FF1'-}
=\frac{\prod_{i=1}^{ p}({ a}_i)_\sigma}{(n)_\sigma\prod_{i=1}^{
s}( b_i)_\sigma} \det A^{m}\det X^{m}{}_{{p+\tilde
p}}{\mathcal{F}}_{s+{\tilde s}+1}\left.\left(a_1+m,\dots ,
a_p+m,{\tilde a}_1,\dots ,{\tilde a}_{\tilde p}+m\atop n+m,
b_1+m,\dots ,b_s+m, {\tilde b}_1, \cdots ,{\tilde b}_{\tilde
s}\right| A, XY\right) \ ,
\end{equation}
\begin{equation}\label{FF1}
\int_{U(n)} \det U^{m}{}_p{\mathcal{F}}_s\left.\left(a_1,\dots
,a_p\atop b_1, \cdots ,b_s\right| A, XU\right){}_{\tilde
p}{F}_{\tilde s}\left.\left({\tilde a}_1,\dots ,{\tilde a}_{\tilde
p}\atop {\tilde b}_1, \cdots ,{\tilde b}_{\tilde s}\right|
U^{-1}Y\right)d_*U
\end{equation}
\begin{equation}\label{FF1'}
=\frac{\prod_{i=1}^{\tilde p}({\tilde
a}_i)_\sigma}{(n)_\sigma\prod_{i=1}^{\tilde s}({\tilde
b}_i)_\sigma} \det Y^{m}{}_{{p+\tilde p}}{\mathcal{F}}_{s+{\tilde
s}+1}\left.\left(a_1,\dots , a_p,{\tilde a}_1+m,\dots ,{\tilde
a}_{\tilde p}+m\atop n+m, b_1,\dots ,b_s, {\tilde b}_1+m, \cdots
,{\tilde b}_{\tilde s}+m\right| A, XY\right)
\end{equation}

\end{consequence}
Let us notice that putting $A=I_n$ in
${}_p{\mathcal{F}}_s(\cdots|A,X)$ one gets ${}_p{F}_s(\cdots|X)$
with the same sets of indices $\{ a_i \},\{ b_i \}$. Choosing
${}_0{F}_0$ of (\ref{examples}) we obtain the simplest example of
(\ref{FF1-})-(\ref{FF1'}):
\begin{equation}\label{Wetting}
\int_{U(n)}\det U^{\mp m}e^{\textrm{Tr}\left( X U
+U^{-1}Y\right)}d_*U= \frac{1}{(n)_\sigma} \left.\left\{\det X^{m}
\atop \det Y^{m}\right.\right\} {}_0F_1(n+m,XY)
\end{equation}
For the determinant representation see (\ref{det1'}). At first
this integral was evaluated in \cite{SW} via methods of
\cite{Bal}. The matrix integral in case $m=0$ was considered in
\cite{Mor} by a different method (by the method of orthogonal
polynomials), and  links with the KP equation and with the
so-called generalized Kontsevich model were established. For $m=0$
and $X=Y=-\frac 1g I_n$ the integral  is used for the study of
two-dimensional QCD \cite{GW} and called Gross-Witten one
plaquette model (about this integral see also \cite{Kos3}):
\begin{equation}\label{GrWi}
I^{GW}(g)=\int_{U(n)}e^{-\frac{1}{g}\textrm{Tr}\left(U
+U^+\right)}d_*U
\end{equation}
In $n \to \infty$ limit the Gross-Witten model enjoys a third
order phase transition \cite{GW}.

The other example of (\ref{FF1-})-(\ref{FF1'}) is ($\sigma$ is the
partition $ (m,\dots , m), l(\sigma)=n$)
\begin{equation}\label{Wetting}
\int_{U(n)}\det U^{\mp m}\det(1- X U)^{-a}\det(1-
U^{-1}Y)^{-{\tilde a}} d_*U= \frac{1} {(n)_\sigma}\left.\left\{({
a})_\sigma\det X^{m} \atop ({\tilde a})_\sigma\det
Y^{m}\right.\right\} {}_2F_1\left.\left(a,{\tilde a}+m \atop
n+m\right | XY\right)
\end{equation}
The determinant representation of the Gauss hypergeometric
function  see (\ref{det1'}). Let us mark that ${}_2F_1(a,b;c|X)$
with integer $a,b,c$ solves Painleve V equation \cite{AvM}.

Let $U$ (and $V$) be $N$ by $N$  (and respectively $n$ by $n$)
unitary matrices, $N\ge n$, and $A,X$ (and $B,Y$)  are $n$ by $N$
(and respectively $N$ by $n$) {\bf rectangle matrices}  . One
combines results of Theorems 3 and 1 to get
\begin{equation}\label{wetting1}
\int_{U(n)}\int_{U(N)} e^{\textrm{Tr}\ XUYV^+ +\textrm{Tr}\
VAU^+B}\  d_*U d_*V ={}_0{\mathcal{F}}_1(n|BX,YA)
\end{equation}
At first the similar result was obtained in \cite{SW} (see also
\cite{Fyod}), where the answer was given as a determinant of
Bessel functions, by (\ref{detF2}) it is the same as
${}_0{\mathcal{F}}_1$ of (\ref{wetting1}). Other examples of
combining of results of Theorems 1,3:
\begin{equation}\label{wetting2}
\int_{U(n)}\int_{U(N)} \det(1-XUYV^+)^{-a} e^{\textrm{Tr}\
VAU^+B}d_*U d_*V={}_1{\mathcal{F}}_1\left.\left(a \atop
n\right|BX,YA \right)
\end{equation}
\begin{equation}\label{wetting3}
\int_{U(n)}\int_{U(N)} \det(1-XUYV^+)^{-a}\det(1-VXU^+Y)^{-b} d_*U
d_*V={}_2{\mathcal{F}}_1\left.\left(a,b \atop n\right|BX,YA
\right)
\end{equation}

\quad

\begin{theorem} Let $z_i,i=1,\dots,n$ be eigenvalues of the matrix $XY$
\begin{equation}\label{ZtauGW}
\int \tau_{ r}\left( n,{\bf t}, XZ \right) \tau_{\tilde r}\left(
n, Z^+Y ,{\bf t}_\infty\right)e^{-\textrm{Tr}ZZ^+}d^2Z
\end{equation}
\begin{equation}\label{ZtauGW'}
= \tau_{r\tilde r}\left( n, {\bf t} ,XY\right)=\frac{
\det\left(z_i^{n-k}\tau_{r\tilde r}(n-k+1,{{\bf t}},{\bf
t}^*(z_i)) \right)_{i,k=1,\dots ,n}}{\Delta({\bf x}^n)}
\end{equation}
 Also
\begin{equation}\label{ZtauGW2}
\int \tau_{ r}\left( n,{\bf t}, X Z\right) \tau_{\tilde r}\left(
n, Z^{+}X^{-1} ,{\bf t}^*\right)e^{-\textrm{Tr}ZZ^+}d^2Z=
\tau_{r_1}\left( n, {\bf t} ,{\bf t}^*\right)\ ,
\end{equation}
where $r_1$ is the following step function:
\begin{equation}\label{Zr1}
r_1(k)=kr(k) {\tilde r}(k)\ ,\ k>0 \ , \quad r(k)=0 \ ,\ k\le 0
\end{equation}
\end{theorem}
The proof follows from (\ref{det1'}). Examples:
\begin{equation}\label{eAZeZ+B}
\int e^{\sum_{m=1}^\infty t_m \textrm{Tr} (XZ)^m} e^{ \textrm{Tr}
(Z^+Y)} e^{-\textrm{Tr}ZZ^+}d^2Z= e^{\sum_{m=1}^\infty t_m
\textrm{Tr} (XY)^m}
\end{equation}
\begin{equation}\label{twoZ}
\int e^{\sum_{m=1}^\infty t_m \textrm{Tr} Z^m}
e^{\sum_{m=1}^\infty t_m^*  \textrm{Tr} (Z^+)^m}
e^{-\textrm{Tr}ZZ^+}d^2Z=  \sum_\lambda (n)_\lambda s_\lambda({\bf
t})s_\lambda({\bf t}^*)
\end{equation}
Let us notice that series in the r.h.s. of (\ref{twoZ}) coincide
with the series for hermitian-antihermitian matrix model and with
the series for the model of normal matrices \cite{HO'}, which in
$n \to \infty$ limit describes the interface dynamics of a spot of
water inside oil \cite{MWZ}.

 Applying
(\ref{taursingular1detX}) and (\ref{taursingular1detXY}) we get
\begin{consequence} Let  $A,X,Y$ are $n$ by $n$ normal
matrices.  For integers $m\ge 0$  we have
\begin{equation}\label{tauZZcon}
\int \tau_{ r}\left( n,A, X Z\right) \tau_{\tilde r}\left( n, Z^+
Y ,{\bf t}_\infty\right)\det(Z^+)^{m}e^{-\textrm{Tr}ZZ^+}d^2Z
\end{equation}
\begin{equation}\label{tauZZcon'}
 =
(n)_{ \sigma}\ r_{ \sigma}(n)\det A^{m}\det X^{m}\
\tau_{r_1}\left( n, A ,XY\right) \ ,
\end{equation}
where the partition $\sigma=(m,\dots , m)$ and are of the length
$l(\sigma)=n$. In (\ref{tauZZcon'})
 $r_1(k)=\frac{k+m}{k}\ r(k+m){\tilde r}(k)$. Also
 \begin{equation}\label{tauZZcon2}
\int \tau_{ r}\left( n,A, X Z\right) \tau_{\tilde r}\left( n, Z^+
Y ,{\bf t}_\infty\right)\det Z^{m}e^{-\textrm{Tr}ZZ^+}d^2Z =
{\tilde r}_{ \sigma}(n)\ \det Y^{m}\ \tau_{r_1}\left( n, A
,XY\right) \ ,
\end{equation}
where the partition $\sigma=(m,\dots , m)$ and are of the length
$l(\sigma)=n$. In (\ref{tauZZcon2})
 $r_1(k)= r(k){\tilde r}(k+m)$.
 In particular
\begin{equation}\label{FF1Z}
\int {}_p{\mathcal{F}}_s\left.\left(a_1,\dots ,a_p\atop b_1,
\cdots ,b_s\right| A,XZ\right){}_{\tilde p}{F}_{\tilde
s}\left.\left({\tilde a_1},\dots ,{\tilde a_{\tilde p}}\atop
{\tilde b_1}, \cdots ,{\tilde b_{\tilde s}}\right|
Z^{+}Y\right)\det (Z^+)^{m}e^{-\textrm{Tr}ZZ^+}d^2Z
\end{equation}
\begin{equation}\label{FF1'Z}
= \frac{\prod_{i=1}^{ p}({ a}_i)_\sigma}{\prod_{i=1}^{
s}({b}_i)_\sigma}\ \det A^{m}\det X^{m}\ {}_{{p+\tilde
p}}{\mathcal{F}}_{s+{\tilde s}}\left.\left(a_1+m,\dots ,
a_p+m,{\tilde a_1},\dots ,{\tilde a_{\tilde p}}\atop b_1+m,\dots
,b_s+m, {\tilde b_1}, \cdots ,{\tilde b_{\tilde s}}\right|
A,XY\right) \ ,
\end{equation}
\begin{equation}\label{FF2Z}
\int {}_p{\mathcal{F}}_s\left.\left(a_1,\dots ,a_p\atop b_1,
\cdots ,b_s\right| A,XZ\right){}_{\tilde p}{F}_{\tilde
s}\left.\left({\tilde a_1},\dots ,{\tilde a_{\tilde p}}\atop
{\tilde b_1}, \cdots ,{\tilde b_{\tilde s}}\right|
Z^{+}Y\right)\det Z^{m}e^{-\textrm{Tr}ZZ^+}d^2Z
\end{equation}
\begin{equation}\label{FF2'Z}
=\frac{\prod_{i=1}^{ {\tilde p}}( {\tilde
a}_i)_\sigma}{\prod_{i=1}^{ s}({\tilde b}_i)_\sigma}\ \det Y^{m}\
{}_{{p+\tilde p}}{\mathcal{F}}_{s+{\tilde s}}\left.\left(a_1,\dots
, a_p,{\tilde a}_1+m,\dots ,{\tilde a}_{\tilde p}\atop b_1,\dots
,b_s, {\tilde b}_1+m, \cdots ,{\tilde b}_{\tilde s}+m\right|
A,XY\right)
\end{equation}
\end{consequence}
For instance
\begin{equation}
\int \det \left(Z^+\right)^m
e^{\textrm{Tr}Z^+Y}e^{\textrm{Tr}XZ-\textrm{Tr}ZZ^+}d^2Z= \det X^m
e^{\textrm{Tr}XY} \ ,
\end{equation}
\begin{equation}
\int \det \left(Z^+\right)^m \det(1-Z^+Y)^{-a}
e^{\textrm{Tr}XZ-\textrm{Tr}ZZ^+}d^2Z= \det X^m\det (1-XY)^{-a} \
,
\end{equation}
\begin{equation}
\int \det \left(Z^+\right)^m \det(1-XZ)^{-a}\det(1-Z^+Y)^{k}
e^{-\textrm{Tr}ZZ^+}d^2Z= (a)_\sigma\det X^m{_2F_0}(-k,a|XY)
\end{equation}

Let $Z_1$ (and $Z_2$) be $N$ by $N$  (and respectively $n$ by $n$)
unitary matrices, $N\ge n$, and $A,X$ (and $B,Y$)  are $n$ by $N$
(and respectively $N$ by $n$) {\bf rectangle matrices}. Theorems
4,2 yields the following examples:
\begin{equation}\label{Zwetting1}
\int_{C^{N^2}} \int_{C^{n^2}}  e^{\textrm{Tr}\ XZ_1YZ_2^+
+\textrm{Tr}\ Z_2AZ_1^+B}e^{-\textrm{Tr}Z_1Z_1^+
-\textrm{Tr}Z_2Z_2^+} d^2Z_1 d^2Z_2 ={}_0{\mathcal{F}}_0(BX,YA)
\end{equation}
\begin{equation}\label{Zwetting2}
\int_{C^{N^2}} \int_{C^{n^2}} \det(1-XZ_1YZ_2^+)^{-a}
e^{\textrm{Tr}\ Z_2AZ_1^+B}e^{-\textrm{Tr}Z_1Z_1^+
-\textrm{Tr}Z_2Z_2^+} d^2Z_1 d^2Z_2
={}_1{\mathcal{F}}_0\left.\left(a \right|BX,YA \right)
\end{equation}

\section{Kontsevich-type integrals}

{\bf Lemma} If $f_1,\dots , f_n$ is a set of functions of one
variable, and $g({\bf x}^n)$ is an antisymmetric function of
$x_1,\dots, x_n$, then
\begin{equation}\label{Kantisymintegral}
\int \cdots \int g({\bf x}^n)\det \left(f_j(x_i)\right)|_{i,j=1}^n
 dx_1\cdots dx_n=n!\int \cdots \int g({\bf x}^n)\prod_{i=1}^n f_i(x_i)  dx_1\cdots
dx_n
\end{equation}
(below we put $g({\bf x}^n)=\Delta({\bf x}^n)$)

Let $X$ be $n$ by $n$ Hermitian matrix with eigenvalues
$x_1,\dots,x_n$, let and $Y$ be $n$ by $n$ matrix with eigenvalues
$y_1,\dots ,y_n$. Let $dX=\prod_{i<k} d\Re X_{ik} d\Im
X_{ik}\prod_{i=1}^n dX_{ii}=d_*U \Delta({\bf x}^n)^2\prod_{i=1}^n
dx_i$. The determinant representation (\ref{det1'}), the lemma and
theorems 1,2 result in

\begin{theorem} The following integrals over Hermitian matrix $X$
and over complex matrix $Z$
\begin{equation}\label{K1}
\int \tau_{r}(n,{\bf t},X)\tau_{\tilde r}(n,XY,I_n)dX= \int
\tau_{r}(n,{\bf t},X)\tau_{\tilde r}(n,ZXZ^+Y,{\bf t}_\infty
)e^{-\textrm{Tr}ZZ^+}d^2ZdX=
\end{equation}
is equal to
\begin{equation}\label{K1'}
\frac{n!C}{\Delta({\bf y}^n)}\int
\det\left(x_i^{n-k}\tau_r(n-k+1,{\bf t}(x_i),{{\bf
t}^*})\tau_{\tilde r}(1,x_i,y_i) \right)_{i,k=1,\dots ,n}
dx_1\cdots dx_n
\end{equation}
\begin{equation}\label{K1''}
=\frac{n!C}{\Delta({\bf y}^n)}\det \left(\int
x^{n-k}\tau_r(n-k+1,{\bf t}(x),{{\bf t}^*})\tau_{\tilde
r}(1,x,y_i) dx \right)_{i,k=1,\dots ,n}
\end{equation}
\end{theorem}
For the proof we use (\ref{GHCIZa}) and the Lemma
(\ref{Kantisymintegral}) and (\ref{det1'}). One obtains the
Kontsevich integral $ \int e^{ t_3\textrm{Tr} X^3}e^{-
\textrm{Tr}XY}dX$ \cite{Ko} if he choose $r= 1$,
$t_k=\delta_{k,3}$ and ${\tilde r}(k)=1/k$.

Also we have
\begin{theorem} Let $a_1,\dots,a_n$ ($y_1,\dots ,y_n$)
are eigenvalues of matrix $A$ (of matrix $Y$)
\begin{equation}\label{K2}
\int \tau_{r}(n,A,X)\tau_{\tilde
r}(n,XY,I_n)dX=\frac{n!C}{\Delta({\bf y}^n)}\det \left(\int
\tau_r(1,a_j,x) \tau_{\tilde r}(1,x,y_i)  dx\right)_{i,k=1,\dots
,n}
\end{equation}
\end{theorem}

\section{New multi-matrix models which can be solved by
the method of orthogonal polynomials and the Schur function
expansion}

Here we consider in short new solvable multi-matrix integrals
which we obtain with the help of the previous consideration,
details will appear in the separate paper.

Let us remind that the  multi-matrix model of Hermitian matrices
\begin{equation}\label{multiold}
I=\int e^{\textrm{Tr}V_1(M_1)+\cdots
+\textrm{Tr}V_N(M_N)}e^{\textrm{Tr}M_1M_2}e^{\textrm{Tr}M_2M_3}\cdots
e^{\textrm{Tr}M_{N-1}M_{N}}  dM_1\cdots dM_N \ ,
\end{equation}
where $M_1,\dots , M_N$ are Hermitian $n$ by $n$ matrices and
\begin{equation}\label{Vk}
  V_k(M_k)=\sum_{m=1}^\infty t^{(k)}_m
M_k^m \ ,\quad k=1,\dots ,N
\end{equation}
was shown to be multi-component KP tau function and solved by the
method of orthogonal polynomials.

Now we consider more general integral over matrices $M_1,\dots ,
M_N$
\begin{equation}\label{multinew}
I=\int e^{\textrm{Tr}V_1(M_1)+\cdots
+\textrm{Tr}V_N(M_N)}R_1(M_1M_2)R_2(M_2M_3)\cdots
R_{N-1}(M_{N-1}M_N) dM_1\cdots dM_N
\end{equation}

 If we choose the interaction term as hypergeometric function of
 matrix argument (see (\ref{taumatr}))
\begin{equation}\label{Rtau}
R_k(M_kM_{k+1})=\tau_{r_k}(n,{\bf t} ,M_kM_{k+1}) \ ,\quad {\bf
t}= (1,0,0,0,\dots)
\end{equation}
then due to (\ref{GHCIZa}) it is possible to perform the angle
integration over each $U_k(n)$ where $M_k=U_kX^{(k)}U_k^+$,\quad $
X^{(k)}=diag(x^{(k)}_1,\dots ,x^{(k)}_n)$. Really, (\ref{GHCIZa})
is a tau function which has a determinant representation
(\ref{det2'}). Finely we obtain the following integral over
eigenvalues $x^{(k)}_i \quad (i=1,\dots ,n; k=1,\dots , N)$:
\begin{equation}\label{eigengenmulti}
I=c\int \prod_{i=1}^n \rho_k(x^{(1)}_ix^{(2)}_i)
\rho_2(x^{(2)}_ix^{(3)}_i) \cdots \rho_{N-1}(x^{(N-1)}_ix^{(N)}_i)
\prod_{k=1}^N e^{\sum_{k=1}^N \sum_{m=1}^\infty  t^{(k)}_m (
x^{(k)}_i )^m } dx^{(k)}_i \ ,
\end{equation}
where
\begin{equation}\label{rhotau}
\rho_{k}(x^{(k)}_ix^{(k+1)}_i)=\tau_{r_k}(1,{\bf
t},x^{(k)}_ix^{(k+1)}_i) \ ,\quad {\bf t}= (1,0,0,0,\dots)
\end{equation}
(In the case of the familiar multi-matrix model (\ref{multiold})
each $\rho_{k}(xy),k=1,\dots,N $ is $e^{-xy} $).

The integral (\ref{eigengenmulti}) may be evaluated by the {\em
method of orthogonal polynomials} $\{p_m( t^{\bf N},x)\}$,
$\{\pi_m(t^{\bf N},x)\}$ which depend on the collection of times
$t^{\bf N}=({\bf t}^{(1)},{\bf t}^{(2)},\dots,{\bf t}^{(N)})$
\begin{equation}\label{orthpol}
\int p_m(t^{\bf N},x)\pi_n(t^{\bf N},y)\omega(t^{\bf N},x,y)
dxdx=\delta_{nm}e^{\phi_n(t^{\bf N})} \ ,\quad n,m=0,1,2,\dots
\end{equation}
Each of polynomials is of a form $p_m({ t^{\bf N}},x)=\sum_{n\le
m}p_{nm}(t^{\bf N}) x^n,\quad \pi_m(t^{\bf N},x)=\sum_{n\le
m}\pi_{nm}(t^{\bf N}) x^n $, and the weight function is
\begin{equation}\label{nu}
\omega(t^{\bf N},x,y)=e^{V_1(x)+V_N(y)}\int \rho_k(xx^{(2)})
\rho_2(x^{(2)}x^{(3)}) \cdots \rho_{N-1}(x^{(N-1)}y)
\prod_{k=2}^{N-1}e^{V_k(x^{(k)})}dx^{(k)}
\end{equation}
We obtain
\begin{equation}\label{tauephi}
I=c\prod_{k=1}^n e^{\phi_k(t^{\bf N})}
\end{equation}
Let us notice that the orthogonal polynomials are Baker-Akhiezer
functions of TL hierarchy. Gauss-Zakharov-Shabat factorization
problem for TL equation  \cite{UT} directly results from the
relation (\ref{orthpol})
\begin{equation}\label{GZS}
K_+(t^{\bf N})=K_-(t^{\bf N})e^\Phi G(t^{\bf N}) \ ,\quad
\Phi=diag(\phi_1(t^{\bf N}),\phi_2(t^{\bf N}),\dots ) \ ,
\end{equation}
where
\begin{equation}\label{KKG}
 (K_+)_{nm}=(\pi^{-1})_{nm} \ ,\quad (K_-)_{nm}=p_{nm} \ ,
 \quad G_{nm}=\int
x^ny^m\omega(t^{\bf N},x,y) dxdx
\end{equation}
(as it was in the case of two-matrix model \cite{GMMMO}).

The Schur function expansion for these matrix matrix models is
given by (\ref{tauschurschur}), where $K_{\lambda\mu}$ are
expressed via products of skew Schur functions, will be presented
in the forthcoming paper.

\section{Acknowledgements}

I thank A.Morozov for a helpful remark and T.Degasperis,
B.Dubrovin, B.Enriquez,  J.Harnad,  J. van de Leur,
M.Mineev-Weinstein, N.Nekrasov and P.Santini for the hospitality
and for discussions and Yu.Neretin for some remarks. I thank IHES
for hospitality and thank grant RFBR 02-02-17382a, LDRD project
20020006ER.


\begin{thebibliography}{99}


\bibitem{Mehta} Mehta,M.L.: {\em Random Matrices}
 Academic Press, Inc, 1991

\bibitem{TW} Tracy,C.A. and Widom,H., Correlation functions,
cluster functions,  and spacing distributions for random matrices,
{\em J. Statist. Phys.} {\bf  92}  809--835 (1998)

\bibitem{ZJZ} Zinn-Justion,P. and Zuber,J.-B.: On some integrals over
the U(N) unitary group and their large limit, {\em
math-ph/0209019}

\bibitem{Moe} van Moerbeke,P.: Integrable
Lattices: Random Matrices and Random Purmutations, {\em
CO/0010135}

\bibitem{Kaz1} Kazakov,V.: Solvable matrix models,
{\em hep-th/003064}

\bibitem{Kaz2} Kazakov,V., Staudacher,M. and Wynter,T.:
 Character Expension Method for Matrix Models of Dually Weighted
 Graphs, {\em hep-th/9502132 v 2}

\bibitem{JM} Jimbo,M and Miwa,T,: Solitons and Infinite
Dimensional Lie Algebras, {\em Publ. RIMS Kyoto Univ.}
 {\bf 19} (1983) 943-1001

\bibitem{M} Morozov,A.Yu.: Integrability and Matrix
 Models, {\em Uspehi Fizicheskih Nauk} {\bf 164} (1994) 3-62

\bibitem{pd22} Orlov,A.Yu. and Scherbin,D.M.: Multivariate
hypergeometric functions as tau functions of Toda lattice and
Kadomtsev-Petviashvili equation, {\em Physics} {\bf D} (2001)

\bibitem{GR} Gross,K.I. and Richards,D.S.: Special functions of matrix
arguments.I: Algebraic induction, zonal polynomials, and
hypergeometric functions, {\em Transactions Amer Math Soc}, {\bf
301} (1987) 781-811

\bibitem{V} Vilenkin,N.Ya. and Klimyk,A.U.:
{\em Representation of Lie Groups and Special Functions.
 Volume 3:
Classical and Quantum Groups and Special Functions}, Kluwer
Academic Publishers, 1992

\bibitem{NTT} Nakatsu,T., Takasaki,K. and Tsujimaru,S.: Quantum and
Classical Aspects of Deformed $c=1$ Strings, {\em Nucl. Phys. B}
{\bf 443} (1995) 155; {\em hep-th/9501038}

\bibitem{T} Takasaki,T.: The Toda Lattice Hierarchy and
Generalized String Equation, {\em Comm.Math.Phys.} {\bf 181}
(1996) 131; {\em hep-th/9506089}

\bibitem{O2} A.Okounkov, Toda equations for Hurwitz numbers,
{\em math.AG/0004128}

\bibitem{Nekrasov} Nekrasov,N.: Seiberg-Witten Prepotential From
Instanton Counting, {\em hep-th/0206161}

\bibitem{1} Orlov,A.Yu.: Soliton theory, Symmetric Functions
and Matrix Integrals, {\em SI/0207030}

\bibitem{HO'} Harnad,J. and Orlov,A.Yu.: Matrix Integrals as
Borel sums of Schur Function Expansions, {\em nlin.SI/0209035}

\bibitem{Or} Orlov,A.Yu.: Tau functions and Matrix
Integrals, {\em math-ph/0210012 v4}

\bibitem{K} Kazakov,V.: {\em Phys Lett B} {\bf 150} (1985) 282

\bibitem{HC} Harish-Chandra:
 {\em Am. J. Math} {\bf 79} (1958)  87-120

\bibitem{IZ} Itzykson,C. and Zuber,J.B.: {\em J.Math.Phys.}
 {\bf 21} (1980) 411

\bibitem{GW} Gross,D.J. and Witten,E.: Possible Third Order
Phase Tranzition in the Large N Lattice Gauge Theory, {\em Phys
Rev D} {\bf 21} (1980)  446

\bibitem{ZSh} Zakharov,V.E. and Shabat,A.B.: {\em J. Funct.
 Anal. Appl.} {\bf 8} (1974)  226

\bibitem{UT} Ueno,K. and Takasaki,K.: {\em Adv. Stud. Pure
Math.} {\bf 4} (1984) 1-95

\bibitem{Mac} Macdonald,I.G.: {\em Symmetric Functions and
Hall Polynomials}, Clarendon Press, Oxford, 1995

\bibitem{Tinit} Takasaki,K.: Initial value problem for the
Toda lattice hierarchy, {\em Adv. Stud. Pure Math.} {\bf 4}
 (1984) 139-163

\bibitem{tmf} Orlov,A.Yu. and Scherbin,D.M.: Hypergeometric solutions
of soliton equations, {\em Theor Math Phys} {\bf 128} (2001)
84-108

\bibitem{HO} Harnad,J. and Orlov,A.Yu.: Scalar products of symmetric
functions and matrix integrals, {\em nlin.SI/0211051} (To appear
in proceedings of NEEDS2002,eds., A. Gonzales,  World Scientific,
2002.)

\bibitem{OP} Okounkov,A and Pandaripande,R.: The equivariant
Gromov-Witten theory of ${\bf P}^1$, {\em math.AG/0207233 v1}

\bibitem{ZJ} Zinn-Justion,P.: HCIZ integral and 2D Toda lattice
hierarchy, {\em math-ph/0202045}

\bibitem{Kos1} Kostov,I.K., Staudacher,M. and Wynter,T.:
Complex Matrix Models and Statistics of Branched Covering of 2D
Surfaces, {\em Commun. Math. Phys.} {\bf 191} (1998) 283-298; {\em
heph-th/9703189}

\bibitem{Kos2} Kostov,I.K.: Exact Solution of the Six-Vertex
Model on a Random Lattice, {\em Nucl. Phys. B} {\bf 575} (2000)
513-534; {\em hep/9911023}

\bibitem{ZKMMO} Zabrodin,A., Kharchev,S., Mironov,A., Marshakov,A.
and Orlov,A.: Matrix Models among Integrable Theories: Forced
Hierarchies and Operator Formalizm, {\em Nuclear Physics B} {\bf
366} (1991) 569-601

\bibitem{SW} Schlittgen,B and Wetting,T: Generalizations of some
integrals over the unitary group", {\em math-ph/0209030 v 1}

\bibitem{Bal} Balantekin,A.B.: Character expensions,
Itzykson-Zuber integrals, and the QCD partition function", {\em
Phys. Rev. D} {\bf 62} (2000) 085017-1\ -\ 085017-8 ;

\bibitem{Mor} Mironov,A., Morozov,A. and Semenoff,G.: Unitary Matrix
Integrals in the Framework of the Generalized Kontsevich Model,
{\em Intern J Mod Phys A} {\bf 11} (1996) 5031-5080

\bibitem{Kos3} Kostov,I.K.: $U(N)$ Gauge Theory and Lattice
Strings, {\em Nucl. Phys. B} {\bf 415} (1994) 29-70; {\em
hep-th/9306110}

\bibitem{AvM} Adler,M. and van Moerbeke,P.: Integrals over
Grassmannians and Random permutations, {\em math.CO/0110281}





\bibitem{Fyod} Fyodorov,Y. and Strahov,E




\bibitem{MWZ} Mineev-Weinstein,M., Wiegmann,P. and Zabrodin,A.:
 Integrable structure of interface dynamics, {\em LAUR-99-0703}

\bibitem{Ko} Kontsevich,M. {\em Funk.Anal. i ego Prilozh.} {\bf 25}
(1991) 50-57

\bibitem{GMMMO}  Gerasimov,A.,  Marshakov,A., Mironov,A.,
Morozov,A. and Orlov,A.: Matrix Models of 2D Gravity and Toda
Theory, {\em Nuclear Physics B} {\bf 357} (1991) 565-618

\end{thebibliography}
\end{document}